\documentclass[epj]{webofc}
\usepackage[utf8]{inputenc}
\usepackage[varg]{txfonts}   % Web of Conferences font
\usepackage{booktabs}
\usepackage{xcolor}
\definecolor{darkred}{rgb}{0.4,0.0,0.0}
\definecolor{darkgreen}{rgb}{0.0,0.4,0.0}
\definecolor{darkblue}{rgb}{0.0,0.0,0.4}
\usepackage[bookmarks,linktocpage,colorlinks,
    linkcolor = darkred,
    urlcolor  = darkblue,
    citecolor = darkgreen]{hyperref}
\wocname{EPJ Web of Conferences}
\woctitle{Lattice2017}
\begin{document}
%%%%%%%%%%%%%%%%%%%%%%%%%%%%%%%%%%%%%%%%%%%%%%%%%%%%%%%%%%%%%%%%%%%%%%%%%%%%%
%
\selectlanguage{english}
%----------------------------------------------------------------------------
\title{%
A precise determination of the HVP contribution to the muon anomalous magnetic moment from lattice QCD
}
%----------------------------------------------------------------------------
\author{%
\firstname{Christoph} \lastname{Lehner}\inst{1}\fnsep\thanks{Speaker, \email{clehner@bnl.gov}}
% \and
%\firstname{Harry} \lastname{Potter}\inst{2}
for the RBC and UKQCD collaborations
}
%----------------------------------------------------------------------------
\institute{%
Physics Department, Brookhaven National Laboratory, Upton, NY 11973, USA
%\and
%Last address unknown
}
%----------------------------------------------------------------------------
\abstract{%
  In this talk I present the current status of a precise
  first-principles calculation of the quark connected, quark
  disconnected, and leading QED and strong isospin-breaking
  contributions to the leading-order hadronic vacuum polarization by
  the RBC and UKQCD collaborations.  The lattice data is also combined
  with experimental $e^+ e^-$ scattering data, consistency between the
  two datasets is checked, and a combined result with smaller error
  than the lattice data and $e^+ e^-$ scattering data individually is
  presented.  }
%----------------------------------------------------------------------------
\maketitle
%----------------------------------------------------------------------------
\section{Introduction}\label{intro}
The anomalous magnetic moment
\begin{align}
  a_\ell = ( g_\ell - 2 ) / 2
\end{align}
of a particle $\ell$ encodes the radiative corrections to Dirac's
result $g_\ell=2$.  These moments are experimentally measured to great
precision for the light leptons $e$ and $\mu$ and therefore provide a
stringent test of the standard model (SM) of particle physics.  Since
contributions of new particles beyond the SM are in general
proportional to the square of the lepton's mass one may expect the
anomalous magnetic moment of the muon to be a particularly interesting
target to find new physics.  Intriguingly, there is a long-standing
tension at the $3\sigma-4\sigma$ level between the SM theory
prediction and the BNL E821 experimental result, see
Tab.~\ref{tab:overview}.

     \begin{table}[hbt]
       \centering
       \begin{tabular}{lrr}\hline\hline
         Contribution & Value $\times 10^{10}$ & Uncertainty $\times 10^{10}$\\\hline
         QED (5 loops) & 11 658 471.895 & 0.008 \\
         EW & 15.4 & 0.1 \\
         HVP LO & 692.3 &  4.2\\
         HVP NLO & -9.84 & 0.06 \\
         HVP NNLO & 1.24 & 0.01 \\
         Hadronic light-by-light & 10.5 &  2.6 \\
         \hline
         Total SM prediction & 11 659 181.5 & 4.9 \\    \hline
         BNL E821 result & 11 659 209.1 & 6.3 \\
         \small FNAL E989/J-PARC E34 goal & & $\approx$ 1.6\\\hline\hline
       \end{tabular}
       \caption{Overview of individual contributions and uncertainties to $a_\mu$.}
       \label{tab:overview}
     \end{table}
     
     With the anticipated fourfold reduction in experimental
     uncertainty targeted by the Fermilab E989 experiment, a similar
     reduction in theory uncertainty is needed.  In this work, we
     address how to reduce the uncertainty in the SM prediction of
     the leading-order (LO) hadronic vacuum polarization (HVP)
     contribution to $a_\mu$.

\section{Lattice methodology}\label{sec-1}
\begin{figure}[htb] % no figure before 1st section
  \centering
  \includegraphics[height=3cm]{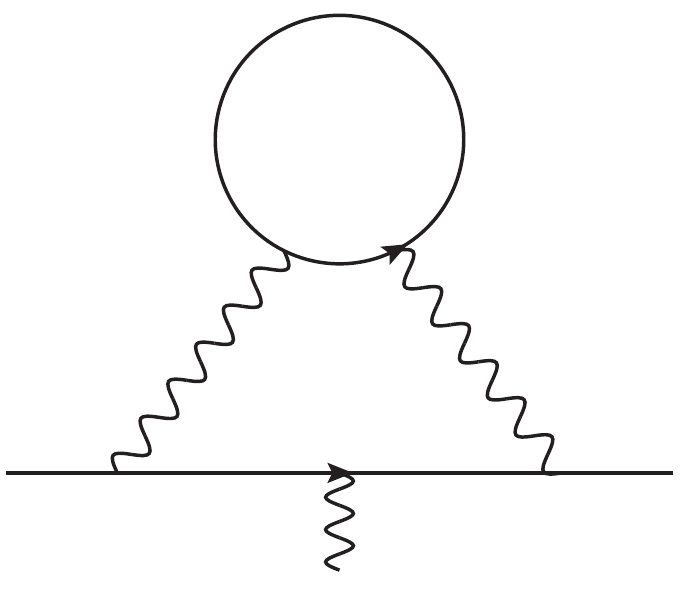} 
  \hspace{1cm}
 \includegraphics[height=3cm]{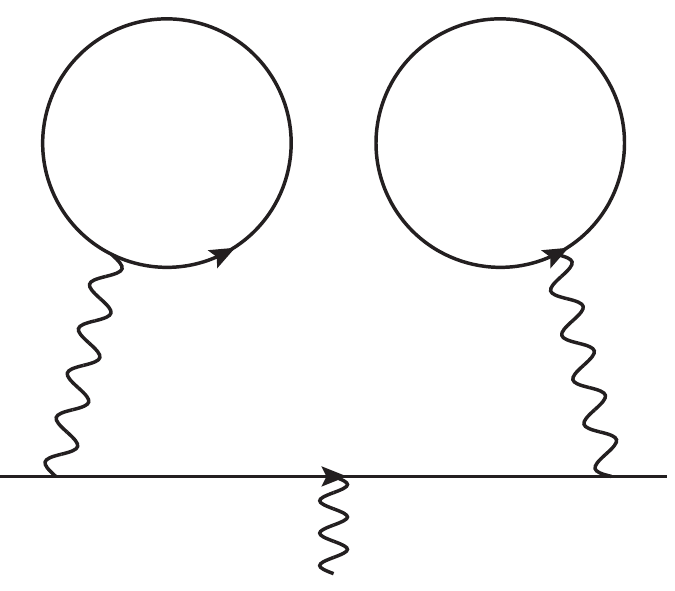} 
  \hspace{1cm}
 \includegraphics[height=3cm]{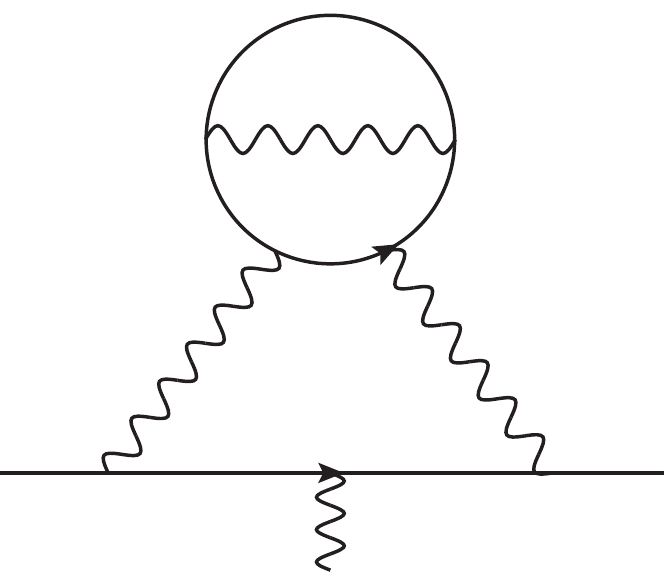} 
   \caption{Quark-connected (left), quark-disconnected (center), and a QED-correction (right) diagram.}
  \label{fig-0}% Give a unique label
\end{figure}

In order to match the target precision of the Fermilab E989
experiment, see Tab.~\ref{tab:overview}, we need to compute
quark-connected, quark-disconnected, and QED and isospin-breaking
contributions to the HVP.  The corresponding diagrams in terms of
quark and photon fields are depicted in Fig.~\ref{fig-0}.  The quark-connected
piece, and in particular the contribution of up and down quarks, is the largest
individual component and needs to be computed at the sub-percent level.
The other contributions are suppressed by at least a factor of ten and therefore
have reduced precision requirements.

In the following sections, we present first-principles lattice QCD
results for all relevant contributions measured on RBC and UKQCD
domain-wall ensembles at effectively physical pion mass.  We use the
$48^3$ and $64^3$ ensembles whose properties are described in detail
in Ref.~\cite{Blum:2014tka}.  We would like to highlight that due to
the precise knowledge of the $\Omega^-$ mass in both experiment and on
the lattice, the lattice spacing is known to 2-3 parts per thousand on
these ensembles.  QED and isospin breaking corrections to this mass
are small and will be presented in an upcoming publication, see
Ref.~\cite{Lehner:2017}.  As was observed recently in
Ref.~\cite{DellaMorte:2017dyu} such high precision on the lattice
spacing is important for a high-precision calculation of $a_\mu$.

It is convenient to discuss the lattice data in the representation of Ref.~\cite{Bernecker:2011gh}
and write
\begin{align}\label{eqn:repr}
  a^{\rm HVP}_\mu &= \sum_{t=0}^\infty w_t C(t) 
\end{align}
with
\begin{align}
  C(t) = \frac13 \sum_{\vec{x}}\sum_{j=0,1,2} \langle J_j(\vec{x},t) J_j(0) \rangle \,,
\end{align}
where
\begin{align}
  J_\mu(x) = i\sum_f Q_f \overline{\Psi}_f(x) \gamma_\mu \Psi_f(x)
\end{align}
is the vector current with sum over all quark flavors $f$ with QED
charge $Q_f$.  The coefficient $w_t$ captures the photon and muon part
of the diagram and can be obtained in a continuum calculation as proposed in
\cite{Blum:2002ii}.  Throughout our lattice calculation we use
appropriately normalized local vector currents.

\subsection{Quark-connected contribution}
\begin{figure}[thb] % no figure before 1st section
  \centering
      \includegraphics[width=9cm]{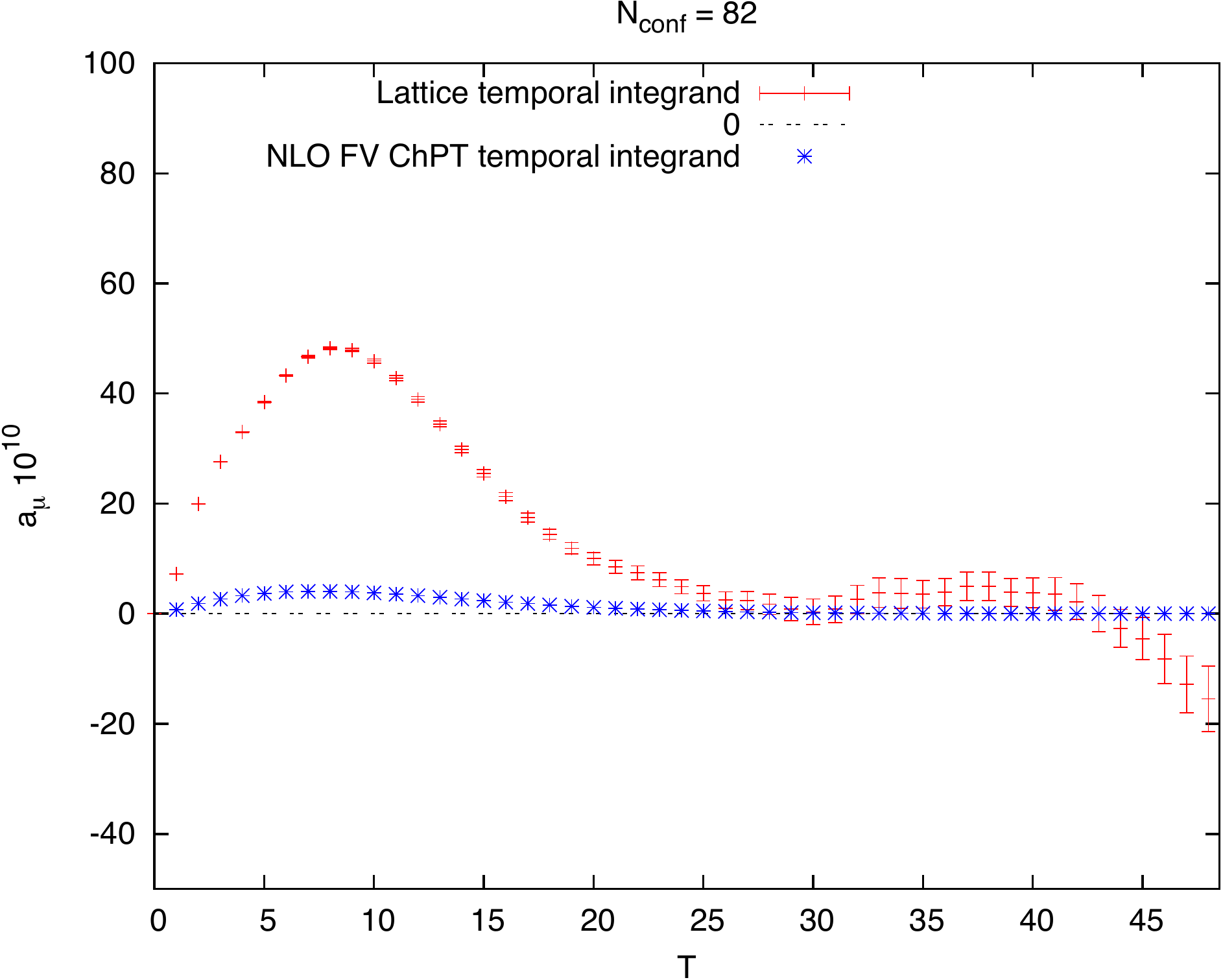}
      \caption{Integrand $w_T C(T)$ for the light-quark connected contribution with $m_\pi=140$ MeV and $a=0.11$ fm.}
  \label{fig-m1}% Give a unique label
\end{figure}

There are two significant obstacles to a high-precision calculation of
the quark-connected contribution to $a_\mu$: the statistical precision
deteriorates exponentially at long distances and the finite-volume
errors may be non-negligible \cite{Aubin:2015rzx}.  In particular the
light-quark contribution suffers from these uncertainties.  In
Fig.~\ref{fig-m1} we show the weighted correlator $w_t C(t)$ and a
corresponding scalar QED calculation in position space that we will
subsequently use to correct leading finite-volume effects.  The
precision degradation at long distances is clearly visible.

\begin{figure}[thb] % no figure before 1st section
  \centering
  \includegraphics[width=7cm]{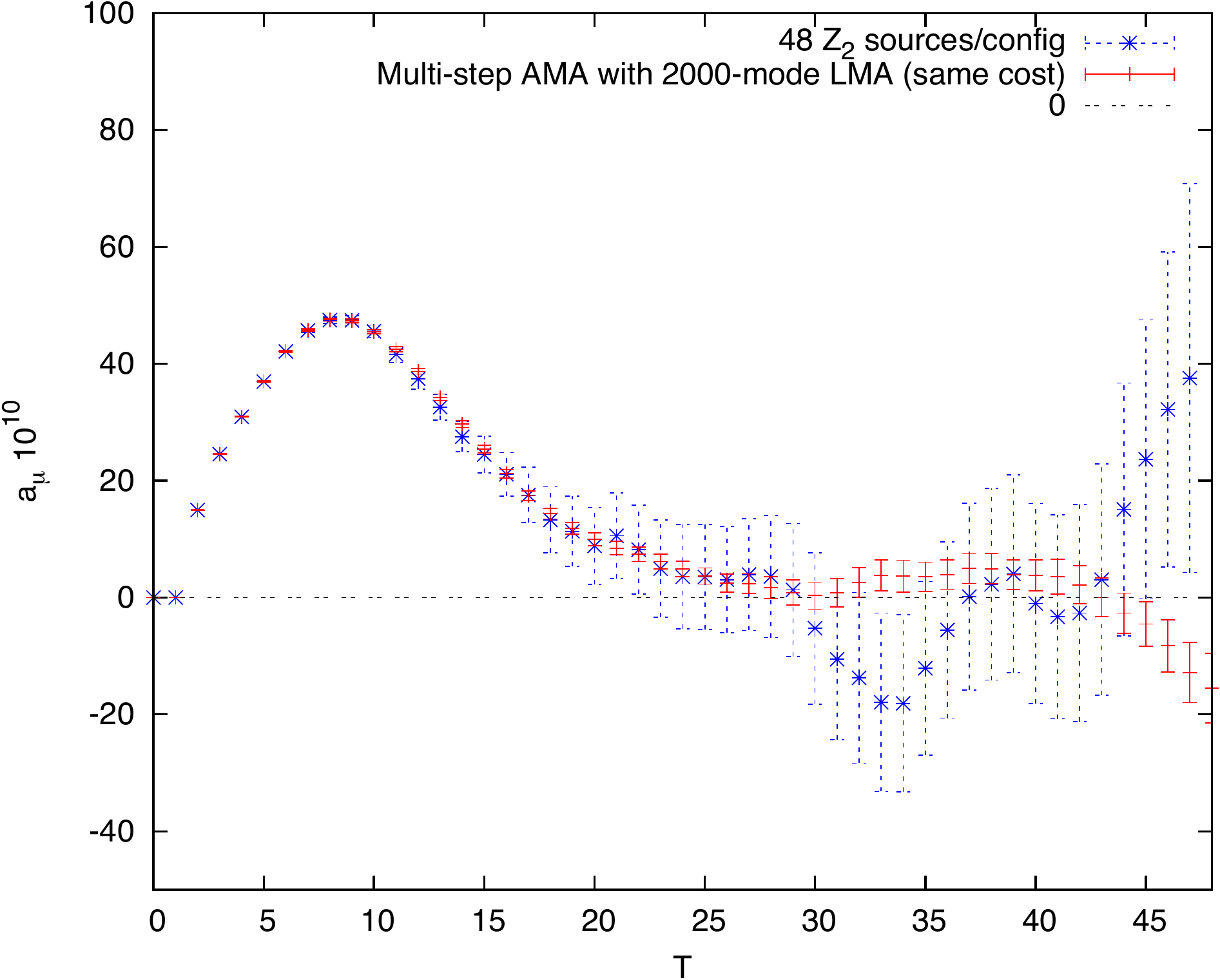}
  \includegraphics[width=7cm]{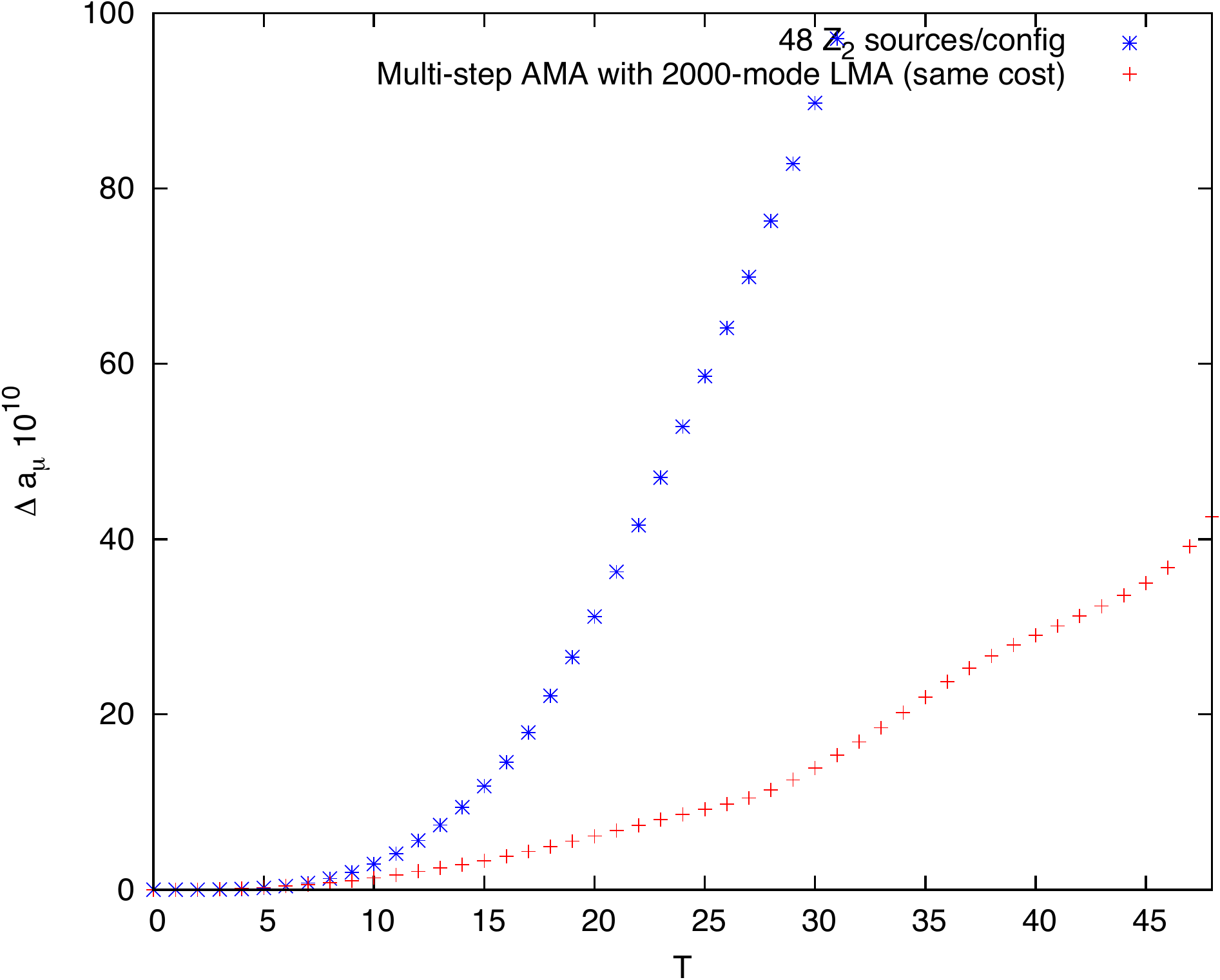}
  \caption{Significant error reduction using full-volume low-mode average  in addition to a multi-level all-mode average.}
%(\cite{\tiny DeGrand \& Sch\"afer 2004})
  \label{fig-m2}% Give a unique label
\end{figure}

We employ a multi-step approximation scheme with low-mode averaging
\cite{DeGrand:2004wh} over the entire volume and two levels of
approximations in a stopped deflated solver (AMA)
\cite{Collins:2007mh,Bali:2009hu,Blum:2012uh,Shintani:2014vja} of
randomly positioned point sources.  Figure~\ref{fig-m2} contrasts the
statistical precision of this improved estimator with the traditional
choice of $Z_2$ wall sources and we observe a reduction of noise by
one order of magnitude at same cost.  We use low modes obtained
through a new Lanczos method working on multiple grids
\cite{Lehner:2017lanc}, which yields an approximate tenfold reduction
in memory cost for our ensembles at physical pion mass with
approximately $5$ fm box size.

For the strange-quark quark-connected contribution we use the dataset published in Ref.~\cite{Blum:2016xpd}.

\subsection{Quark-disconnected contribution}
Our result for the quark-disconnected diagram at physical pion mass is published in Ref.~\cite{Blum:2015you},
where we find
\begin{align}
  a_\mu^{\rm HVP~(LO)~DISC} = -9.6(3.3)_{\rm stat}(2.3)_{\rm sys} \times 10^{-10} \,.
\end{align}

\subsection{QED and strong isospin-breaking corrections}
\begin{figure}[thb] % no figure before 1st section
  \centering
  \includegraphics[width=9cm]{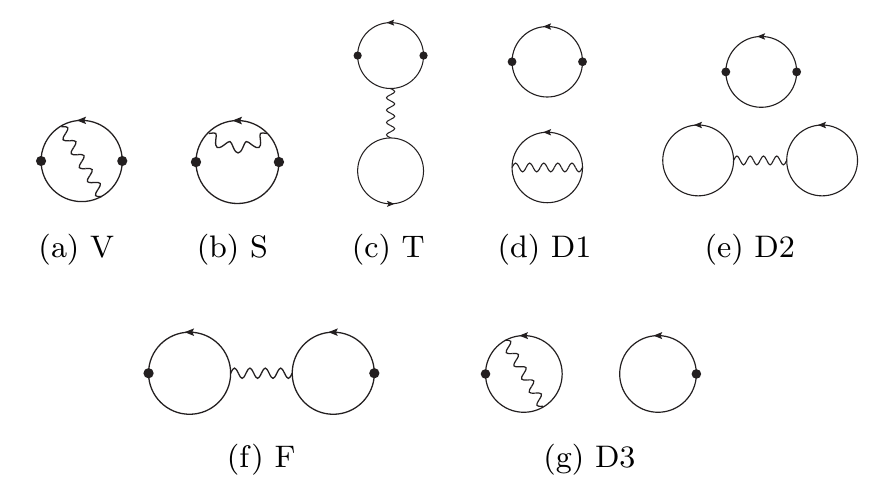}
  \caption{QED correction diagrams.}
%(\cite{\tiny DeGrand \& Sch\"afer 2004})
  \label{fig-m3}% Give a unique label
\end{figure}

\begin{figure}[thb] % no figure before 1st section
  \centering
  \includegraphics[width=7cm]{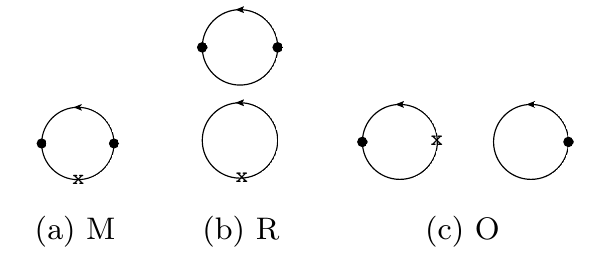}
  \caption{Strong isospin-breaking correction diagrams.  The crosses
    denote a sequential insertion of a scalar operator.}
%(\cite{\tiny DeGrand \& Sch\"afer 2004})
  \label{fig-m3b}% Give a unique label
\end{figure}

We compute both QED corrections and strong isospin-breaking
corrections to our result, see Figs.~\ref{fig-m3} and \ref{fig-m3b},
respectively.  We utilize an importance sampling technique in position
space similar to Ref.~\cite{Blum:2015gfa} and an appropriately
normalized local vector current coupling to the photon fields.  The
photons are regulated with the QED$_L$ prescription in this work and
the universal $1/L$ and $1/L^2$ corrections are taken into account for
the mass spectrum \cite{Borsanyi:2014jba}.  The use of an
infinite-volume photon to reduce finite-volume effects in the QED
corrections to the HVP is currently also being explored and will be
presented in Ref.~\cite{Lehner:2017}.  The results presented in
this section are at physical pion mass improving over our recent work
\cite{Boyle:2017gzv}.

We first compute QED and strong isospin-breaking corrections to the
$K^0$, $K^+$, $\pi^0$, $\pi^+$, and $\Omega^-$ masses and re-tune the
up, down, and strange quark mass as well as the lattice spacing by
matching these masses to their experimental values.

The re-tuned quark masses and the lattice spacing can be expanded
around the values of the pure QCD simulation with light quark mass $m_l$,
strange quark mass $m_h$, and lattice spacing $a^{(0)}$.  We write
\begin{align}
  m_u &= m_l + \Delta m_u \,, \\
  m_d &= m_l + \Delta m_d \,, \\
  m_s &= m_h + \Delta m_s \,, \\
  a   &= a^{(0)} + a^{(1)}\,.
\end{align}

The pion and kaon spectrum provides the following conditions
\begin{align}
a m^{(exp)}_{\pi^+} &= m^{(0)}_{\pi} + m^{(1,\rm QED)}_{\pi^+} + m^{(1,u)}_{\pi^+} \Delta m_u  + m^{(1,d)}_{\pi^+} \Delta m_d \,, \\
a m^{(exp)}_{\pi^0} &= m^{(0)}_{\pi} + m^{(1,\rm QED)}_{\pi^0} + m^{(1,u)}_{\pi^0} \Delta m_u  + m^{(1,d)}_{\pi^0} \Delta m_d \,, \\
a m^{(exp)}_{K^+} &= m^{(0)}_{K} + m^{(1,\rm QED)}_{K^+} + m^{(1,u)}_{K^+} \Delta m_u  + m^{(1,s)}_{K^+} \Delta m_s \,, \\
a m^{(exp)}_{K^0} &= m^{(0)}_{K} + m^{(1,\rm QED)}_{K^0} + m^{(1,d)}_{K^0} \Delta m_d  + m^{(1,s)}_{K^0} \Delta m_s \,,
\end{align}
where only leading QED ($m^{(1,\rm QED)}$) and mass-correction ($m^{(1,u/d/s)}$) effects have been included.
The effect of the lattice-spacing shift can be made more explicit in
\begin{align}
  a^{(0)} m^{(exp)}_{\pi^+} &= m^{(0)}_{\pi} + m^{(1,\rm QED)}_{\pi^+} + m^{(1,u)}_{\pi^+} \Delta m_u  + m^{(1,d)}_{\pi^+} \Delta m_d  
  - a^{(1)} m^{(exp)}_{\pi} \,, \\
  a^{(0)} m^{(exp)}_{\pi^0} &= m^{(0)}_{\pi} + m^{(1,\rm QED)}_{\pi^0} + m^{(1,u)}_{\pi^0} \Delta m_u  + m^{(1,d)}_{\pi^0} \Delta m_d 
  - a^{(1)} m^{(exp)}_{\pi} \,, \\
  a^{(0)} m^{(exp)}_{K^+} &= m^{(0)}_{K} + m^{(1,\rm QED)}_{K^+} + m^{(1,u)}_{K^+} \Delta m_u  + m^{(1,s)}_{K^+} \Delta m_s 
  - a^{(1)} m^{(exp)}_{K} \,, \\
  a^{(0)} m^{(exp)}_{K^0} &= m^{(0)}_{K} + m^{(1,\rm QED)}_{K^0} + m^{(1,d)}_{K^0} \Delta m_d  + m^{(1,s)}_{K^0} \Delta m_s 
  - a^{(1)} m^{(exp)}_{K} \,.
\end{align}

If we restrict the discussion of mass-shifts to diagram $M$, a further simplification
can be obtained
\begin{align}
  a^{(0)} m^{(exp)}_{\pi^+} &= m^{(0)}_{\pi} + m^{(1,\rm QED)}_{\pi^+} + m^{(1,u)}_{\pi^+} (\Delta m_u  + \Delta m_d )
  - a^{(1)} m^{(exp)}_{\pi} \,, \\
  a^{(0)} m^{(exp)}_{\pi^0} &= m^{(0)}_{\pi} + m^{(1,\rm QED)}_{\pi^0} + m^{(1,u)}_{\pi^+} (\Delta m_u  + \Delta m_d )
  - a^{(1)} m^{(exp)}_{\pi} \,, \\
  a^{(0)} m^{(exp)}_{K^+} &= m^{(0)}_{K} + m^{(1,\rm QED)}_{K^+} + m^{(1,u)}_{K^+} \Delta m_u  + m^{(1,s)}_{K^+} \Delta m_s 
  - a^{(1)} m^{(exp)}_{K} \,, \\
  a^{(0)} m^{(exp)}_{K^0} &= m^{(0)}_{K} + m^{(1,\rm QED)}_{K^0} + m^{(1,d)}_{K^0} \Delta m_d  + m^{(1,s)}_{K^0} \Delta m_s 
  - a^{(1)} m^{(exp)}_{K} \,,
\end{align}
which makes it apparent that $a^{(1)}$ and $\Delta m_u + \Delta m_d$
cannot be independently obtained from the above set of equations.  The
expected smaller size of diagram O compared to diagram M likely makes
it at least numerically challenging to obtain a precise determination
of $a^{(1)}$ as well as the mass shifts from the above set of
equations.  We are therefore encouraged to use an alternative mass
scale to determine $a^{(1)}$ for which we will use the mass of the
$\Omega^-$.

If we ignore the lattice-spacing shift for now and only focus on diagrams $V$, $S$, and $M$, we obtain
\begin{align}
\Delta m_u &= -0.000678(83) \,, \\
\Delta m_d &= 0.000519(83) \,, \\
\Delta m_s &= -0.000431(32) \,.
\end{align}
The re-tuning of the lattice spacing is a small effect that is absent
in this work, however, will be presented separately in
Ref.~\cite{Lehner:2017}.

\begin{figure}[thb] % no figure before 1st section
  \centering
  \includegraphics[width=8cm,page=4]{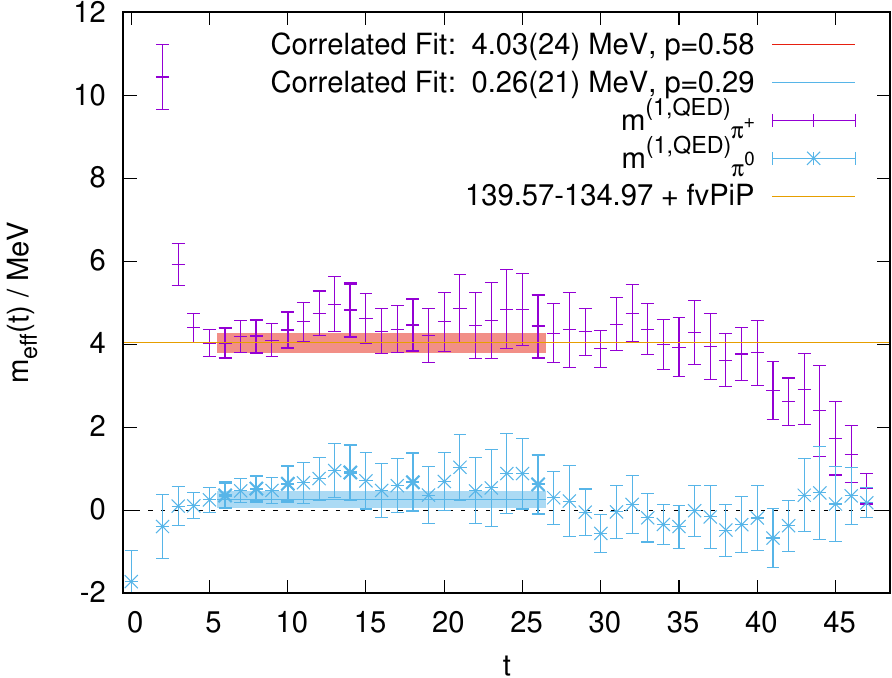}
  \caption{Pion mass splitting at physical pion mass from the
    RBC/UKQCD $48^3$ ensemble.  The lattice data shown here is not yet
    finite-volume corrected, however, to facilitate a proper
    comparison we also show the result of adding the universal QED$_L$
    finite-volume correction (fvPiP) to the experimental pion mass
    splitting.}
%(\cite{\tiny DeGrand \& Sch\"afer 2004})
  \label{fig-m4}% Give a unique label
\end{figure}

Figure~\ref{fig-m4} displays the pion mass splitting taking into account the finite-volume
corrections and shows agreement with the experimental measurement.

We are now in a position to compute the QED and strong
isospin-breaking corrections to $a_\mu^{\rm HVP~LO}$.  In this work,
results for the electro-quenched diagrams $S$, $V$, and $F$ are shown.
 In addition to the absence of a
re-tuning of the lattice spacing, also the effects of QED corrections
on the local-current renormalization are ignored in this work.  These
effects are small and will be presented elsewhere \cite{Lehner:2017}.

\begin{figure}[tbp] % no figure before 1st section
  \centering
  \includegraphics[width=8cm,page=20]{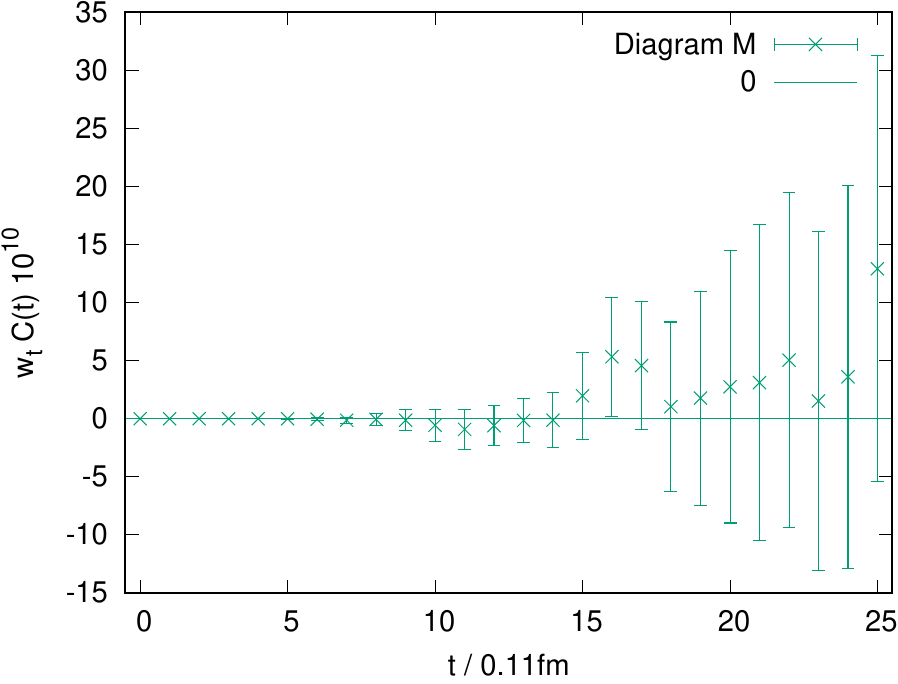}
  \caption{Valence strong isospin-breaking correction to $a_\mu^{\rm HVP~LO}$.}
%(\cite{\tiny DeGrand \& Sch\"afer 2004})
  \label{fig-m5}% Give a unique label
\end{figure}

\begin{figure}[tbp] % no figure before 1st section
  \centering
  \includegraphics[width=8cm,page=23]{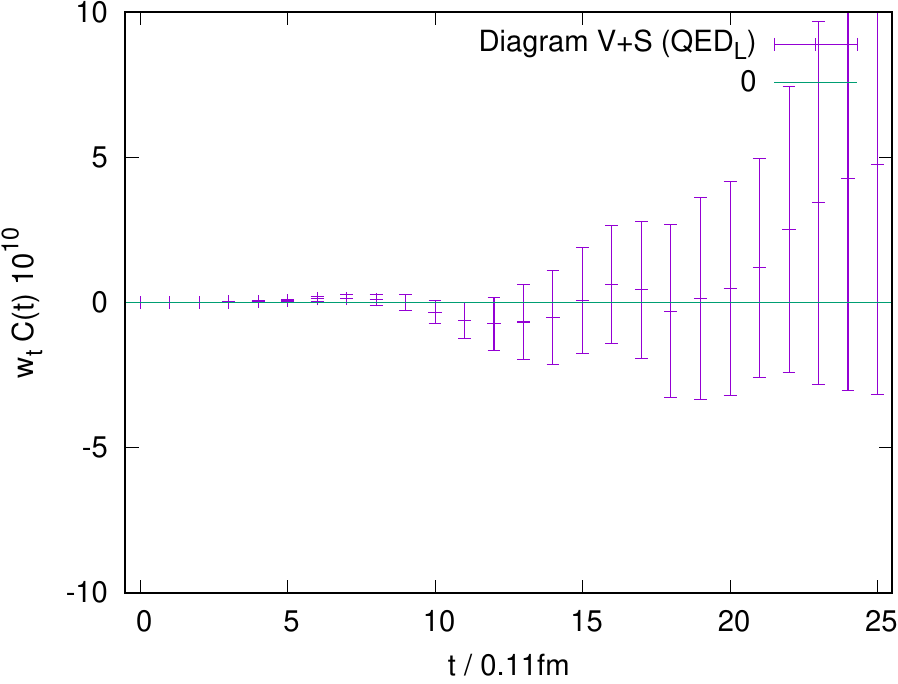}
  \caption{QED corrections from diagrams S and V to $a_\mu^{\rm HVP~LO}$.}
%(\cite{\tiny DeGrand \& Sch\"afer 2004})
  \label{fig-m6}% Give a unique label
\end{figure}

The valence strong isospin-breaking effect (diagram M) is shown in Fig.~\ref{fig-m5}
and the diagrams S and V are shown in Fig.~\ref{fig-m6}.

\section{A combined analysis of lattice data and R-ratio data}
The lattice data presented so far is sufficient to attempt a
comparison to the $e^+ e^-$ scattering data.
We connect $C(t)$ to the R-ratio data \cite{Bernecker:2011gh} writing
\begin{align}
  \Pi(-Q^2) &=\frac{1}{\pi} \int_0^\infty ds \frac{s}{s+Q^2}  \sigma(s,e^+e^- \to {\rm had})
\end{align}
with
\begin{align}
  R(s) = \frac{\sigma(s,e^+e^- \to {\rm had})}{\sigma(s,e^+ e^- \to \mu^+ \mu^-, {\rm tree})}
 = \frac{3 s}{4 \pi \alpha^2} \sigma(s,e^+e^- \to {\rm had}) \,.
\end{align}

\begin{figure}[tbp] % no figure before 1st section
  \centering
      \includegraphics[width=7cm,page=24]{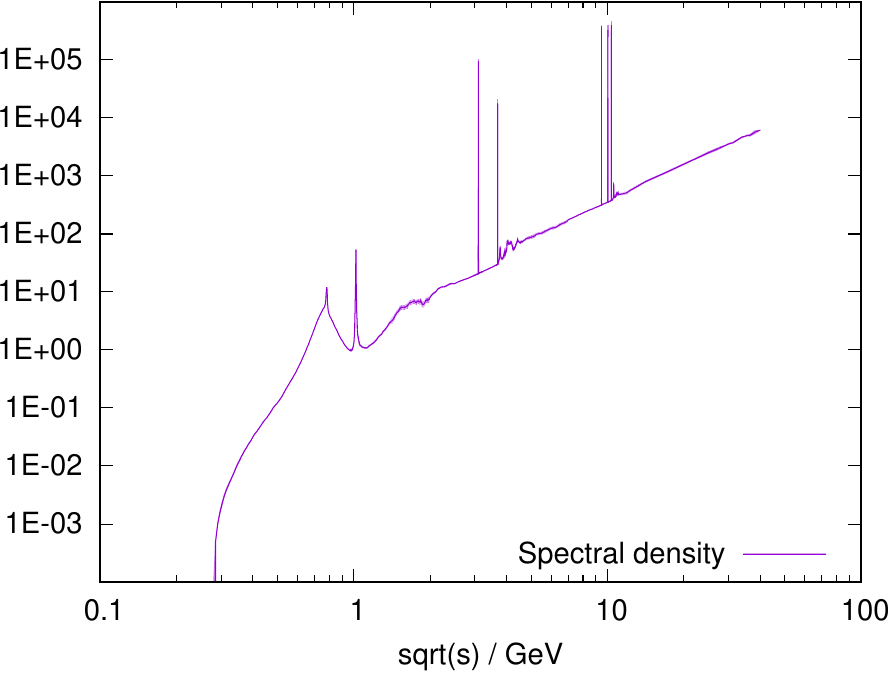}
  \caption{Spectral density function in the convention of Eq.~\eqref{eqn:spec}.}
%(\cite{\tiny DeGrand \& Sch\"afer 2004})
  \label{fig-m8}% Give a unique label
\end{figure}

A Fourier transform then gives
\begin{align}\label{eqn:spec}
  C(t) \propto  \int_0^\infty d(\sqrt{s}) R(s) s e^{-\sqrt{s} t} \equiv \int_0^\infty d(\sqrt{s}) \rho(\sqrt{s}) e^{-\sqrt{s} t} 
\end{align}
which defines a density function $\rho(\sqrt{s})$.  In the remainder of this presentation,
we take $R(s)$ from Ref.~\cite{alphaQED16} and for now treat values for different $s$ to be fully
correlated.  This imperfection is removed in Ref.~\cite{Lehner:2017}.  Figure~\ref{fig-m8}
shows the spectral density function in the convention of Eq.~\eqref{eqn:spec}.

\begin{figure}[bp] % no figure before 1st section
  \centering
      \includegraphics[width=7cm,page=4]{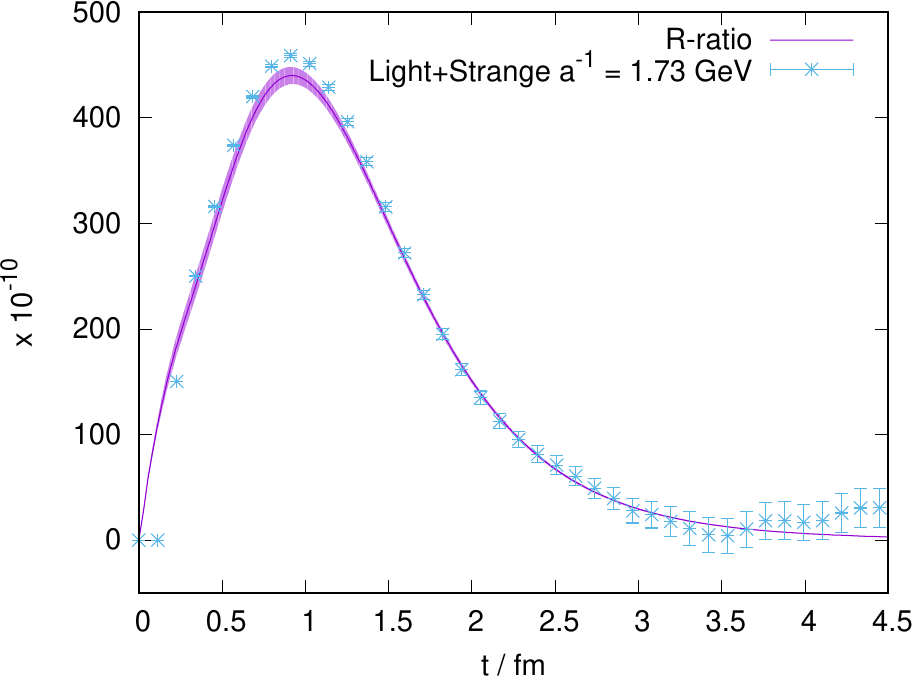}
      \includegraphics[width=7cm,page=5]{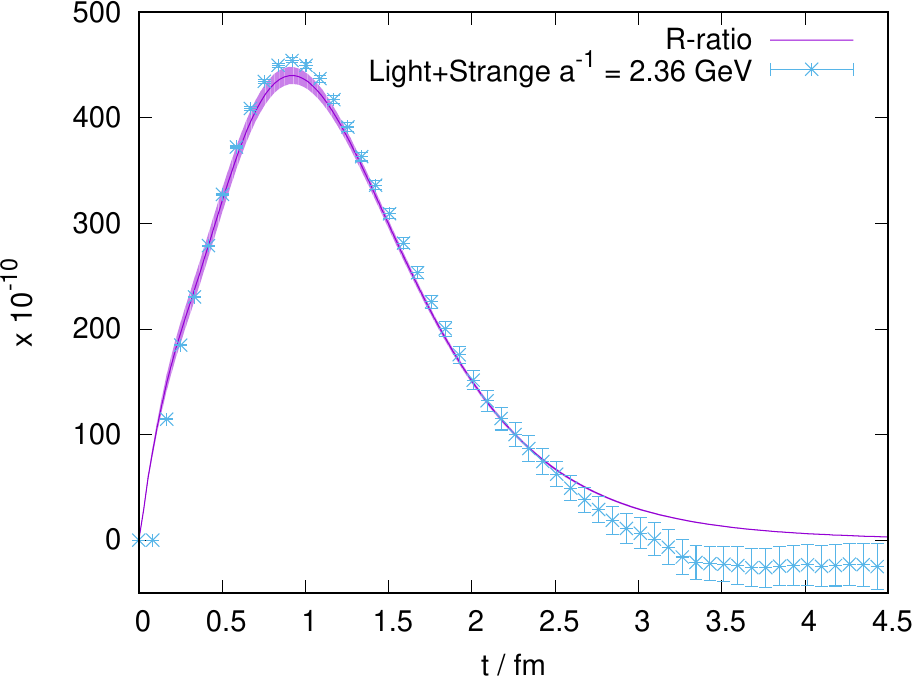}
  \caption{Comparison of R-ratio and lattice data.}
%(\cite{\tiny DeGrand \& Sch\"afer 2004})
  \label{fig-m9}% Give a unique label
\end{figure}

In Fig.~\ref{fig-m9} we contrast $a_\mu$ in the representation of
Eq.~\eqref{eqn:repr} from the R-ratio data with lattice data in the
isospin symmetric limit at two different lattice spacings.  We observe
reasonable agreement and notice that while the lattice
data is most precise at short distances, the R-ratio data behaves in a
complementary manner.

\subsection{Window method}
We define a window method similar to Ref.~\cite{Bernecker:2011gh}
with the intention to combine the regions of respective high precision from
the R-ratio and lattice data sets.
We select a window in $t$ by defining a smeared $\Theta$ function
\begin{align}
\Theta(t,\mu,\sigma) \equiv \left[1 + \tanh\left[ (t-\mu) / \sigma \right]\right]/2
\end{align}
which we find to be helpful to control the effect of discretization errors by the smearing
parameter $\sigma$.

We define the ``window method'' as
\begin{align}
  a_\mu = \sum_t w_t C(t) \equiv a_\mu^{\rm SD} + a_\mu^{\rm W} + a_\mu^{\rm LD} 
\end{align}
with
\begin{align}
  a_\mu^{\rm SD} &= \sum_t C(t) w_t [1 - \Theta(t,t_0,\Delta)] \,, \\
  a_\mu^{\rm W} &= \sum_t C(t) w_t [ \Theta(t,t_0,\Delta) - \Theta(t,t_1,\Delta) ] \,, \\
  a_\mu^{\rm LD} &= \sum_t C(t) w_t \Theta(t,t_1,\Delta)\,,
\end{align}
where each contribution is accessible from both lattice and R-ratio data.

We then take both the short-distance and long-distance region from the R-ratio
data aided by perturbative QCD \cite{alphaQED16}.  The removal of the short-distance
region reduces discretization errors in our calculation, however, is not strictly
required to achieve high precision.

\begin{figure}[tb] % no figure before 1st section
    \begin{center}
      \includegraphics[width=7cm,page=3]{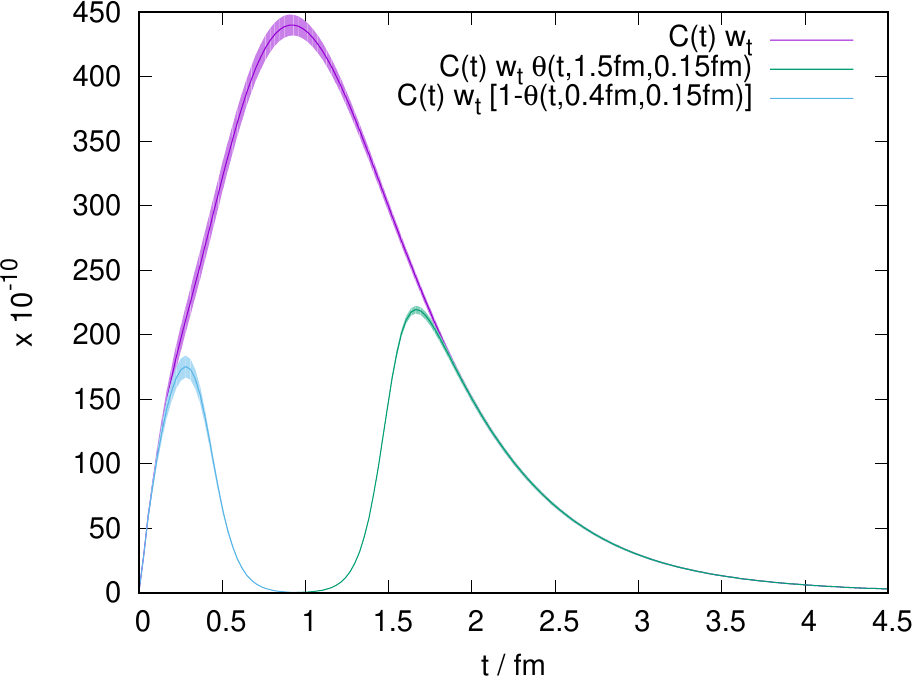}
      \includegraphics[width=7cm,page=25]{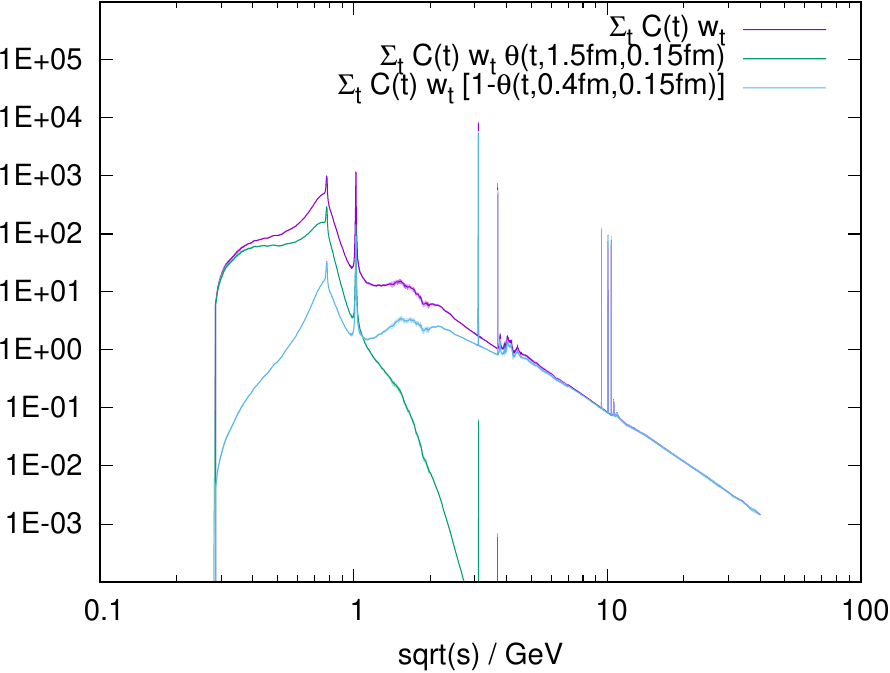}
    \end{center}
  \caption{Window of R-ratio data in Euclidean position space (left) and the effect of the window in terms of re-weighting energy regions (right).}
%(\cite{\tiny DeGrand \& Sch\"afer 2004})
  \label{fig-m9a}% Give a unique label
\end{figure}

In Fig.~\ref{fig-m9a} we show the effect of the window selection in
Euclidean position space and the corresponding effect in terms of
re-weighting different energy regions in their contribution
to $a_\mu$.  The suppression of short distances has a pronounced
effect of high-energy contributions as expected.  In addition the
contribution for long Euclidean distances comes mostly from the lower
end of the energy spectrum.

\begin{figure}[ptb] % no figure before 1st section
  \begin{center}
    \includegraphics[width=8cm,page=6]{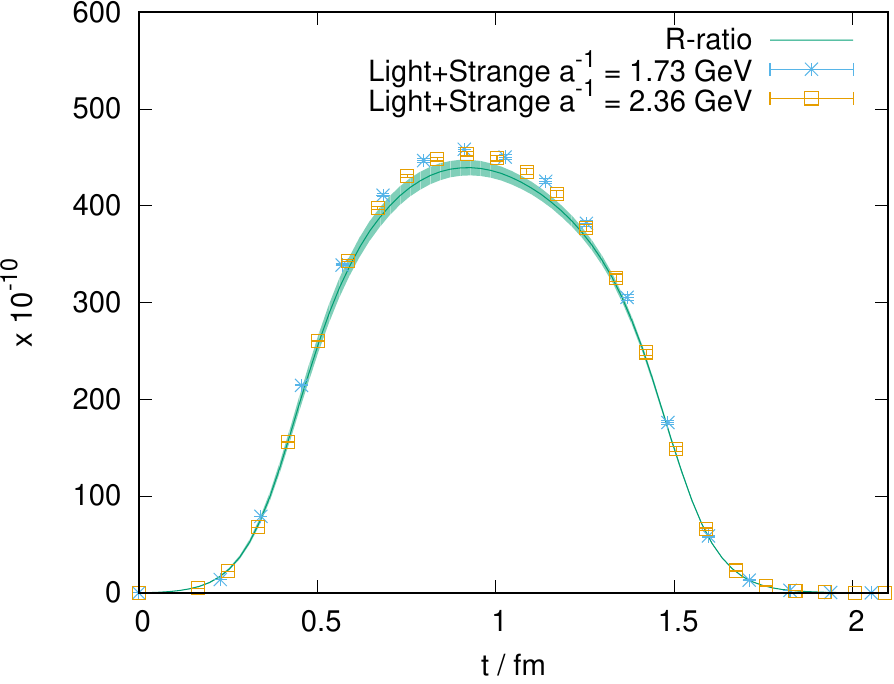}
  \end{center}
  \caption{    Example contribution to $a_\mu^{\rm W}$ with $t_0=0.4$ fm, $t_1=1.5$ fm, $\Delta=0.15$ fm.}
  % (\cite{\tiny DeGrand \& Sch\"afer 2004})
  \label{fig-m9c}% Give a unique label
\end{figure}

Finally Fig.~\ref{fig-m9c} contrasts lattice and R-ratio data for the
window with $t_0=0.4$ fm, $t_1=1.5$ fm, and $\Delta=0.15$ fm.  We
observe good agreement between both datasets.

\subsection{Continuum limit}
\begin{figure}[ptb] % no figure before 1st section
  \begin{center}
    \includegraphics[width=8cm,page=9]{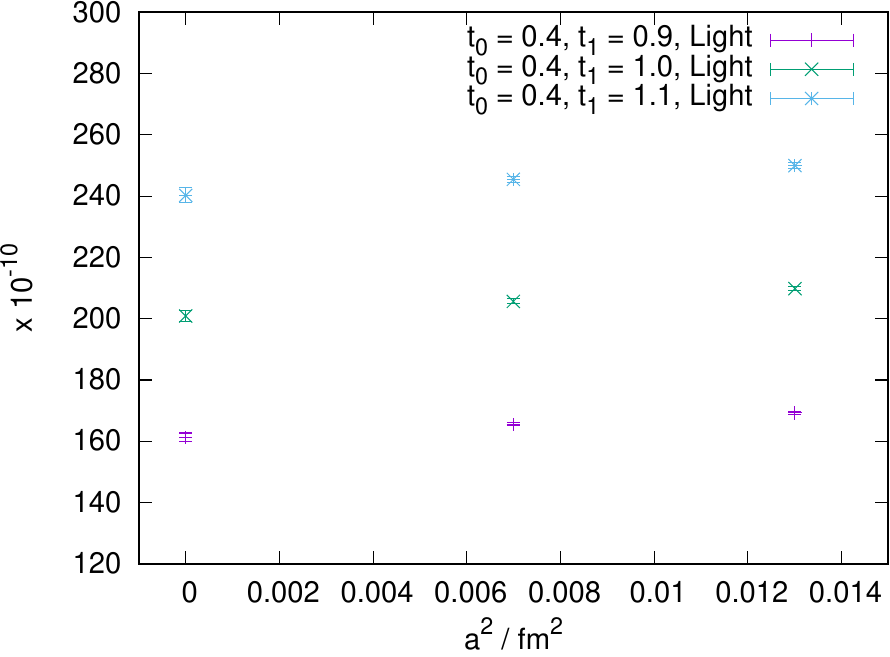}
  \end{center}
  \caption{Continuum limit of $a_\mu^{\rm W}$ with $t_0=0.4$ fm and $\Delta=0.15$ fm.}
  % (\cite{\tiny DeGrand \& Sch\"afer 2004})
  \label{fig-m9d}% Give a unique label
\end{figure}

The continuum limit of our domain-wall fermion lattice calculations at
physical pion mass is mild.  This is illustrated in Fig.~\ref{fig-m9d}
for the light-quark contribution to $a_\mu^{\rm W}$ with $t_0=0.4$ fm
and $\Delta=0.15$ fm, where our coarse-lattice result only differs
from the continuum extrapolation at the level of a few per-cent.  This
is in contrast to the rather steep continuum extrapolation one
observes using staggered fermions, see,
e.g.~Ref.~\cite{Chakraborty:2016mwy}.

For the small quark-disconnected contribution and the QED and strong
isospin-breaking corrections we currently only use one lattice cutoff
($a^{-1}=1.73$ GeV) and include an appropriate estimate of cutoff
effects in the final analysis.

\subsection{Combined analysis}
As explained above, we take $a_\mu^{\rm LD}$ and $a_\mu^{\rm SD}$ from
the R-ratio data and combine it with $a_\mu^{\rm W}$ corresponding to
the continuum and infinite-volume limit of the quark-connected,
quark-disconnected, and QED and strong isospin-breaking contributions
that we have discussed previously.  The contributions of up, down, strange,
and charm quarks are included.

\begin{figure}[ptb] % no figure before 1st section
      \begin{center}
        \includegraphics[width=7cm,page=12]{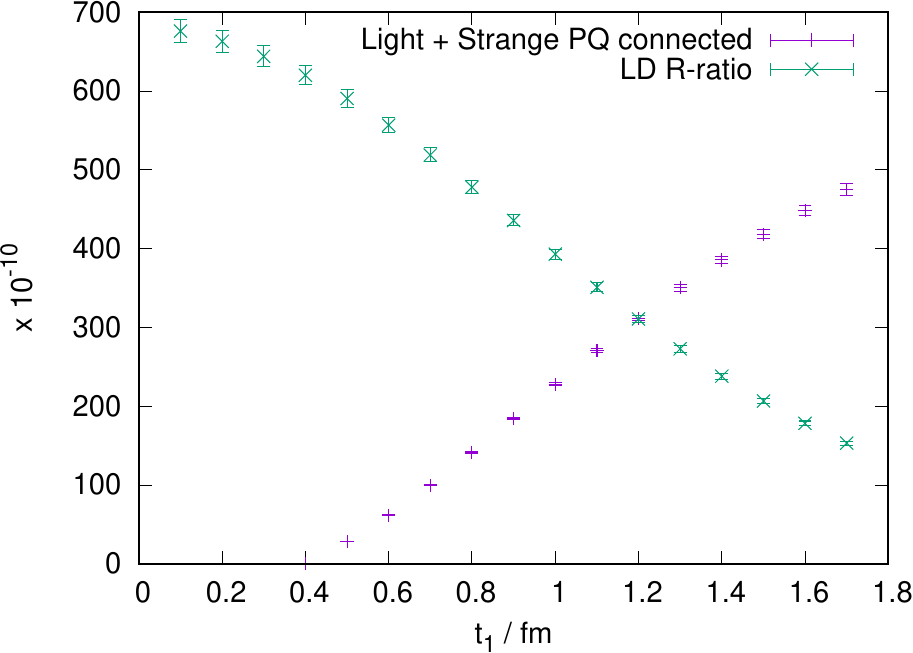}
        \includegraphics[width=7cm,page=13]{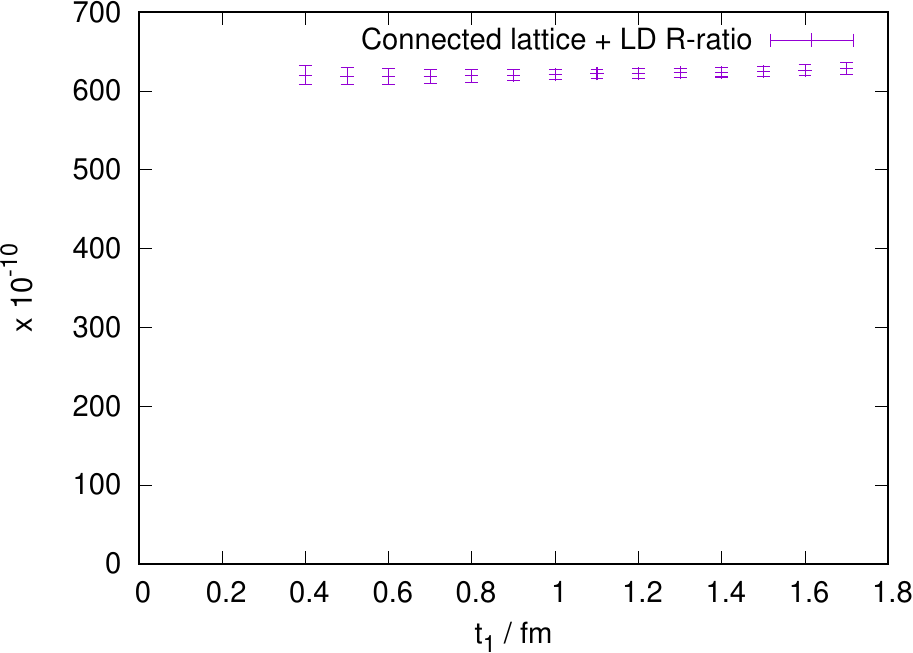}
        \vspace{-0.5cm}
      \end{center}
      \caption{Lattice and R-ratio data contributions seperate (left)
        and summed (right) as a function of $t_1$.  We notice that
        while the individual contributions vary on the order of the
        total $a_\mu$ in this range of $t_1$, the sum is constant
        within its uncertainty.  This provides a strong consistency
        check between the lattice and the R-ratio data.}
      % (\cite{\tiny DeGrand \& Sch\"afer 2004})
      \label{fig-m9e}% Give a unique label
    \end{figure}

\begin{figure}[ptb] % no figure before 1st section
      %\begin{center}
        \begin{minipage}{7cm}
          \begin{center}
            \includegraphics[width=5.5cm,page=14]{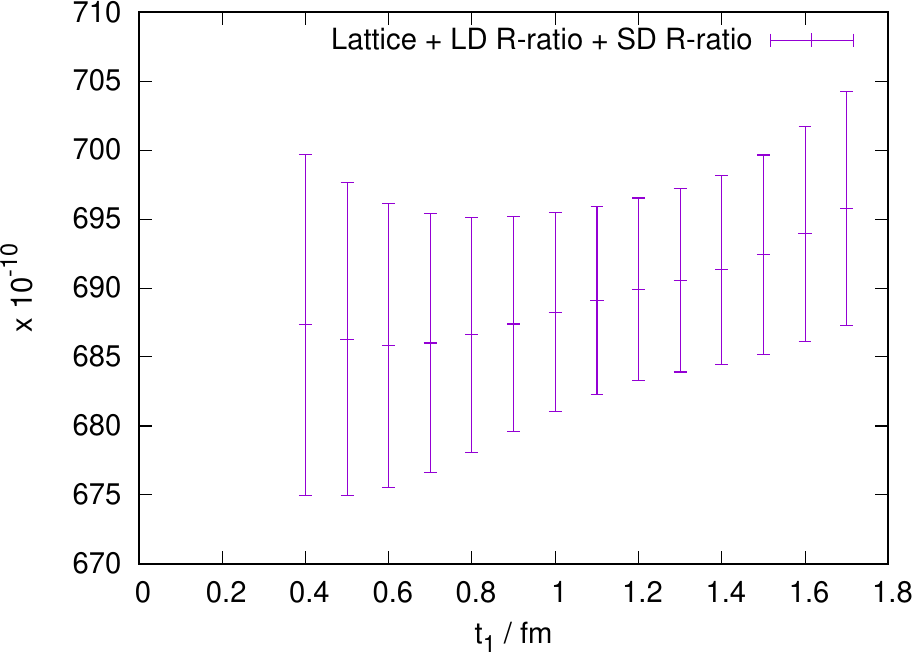}\\
            {\footnotesize Identical to Fig.~\ref{fig-m9e}}
          \end{center}
        \end{minipage}
        \begin{minipage}{7cm}
          \begin{center}
            \includegraphics[width=5.5cm,page=15]{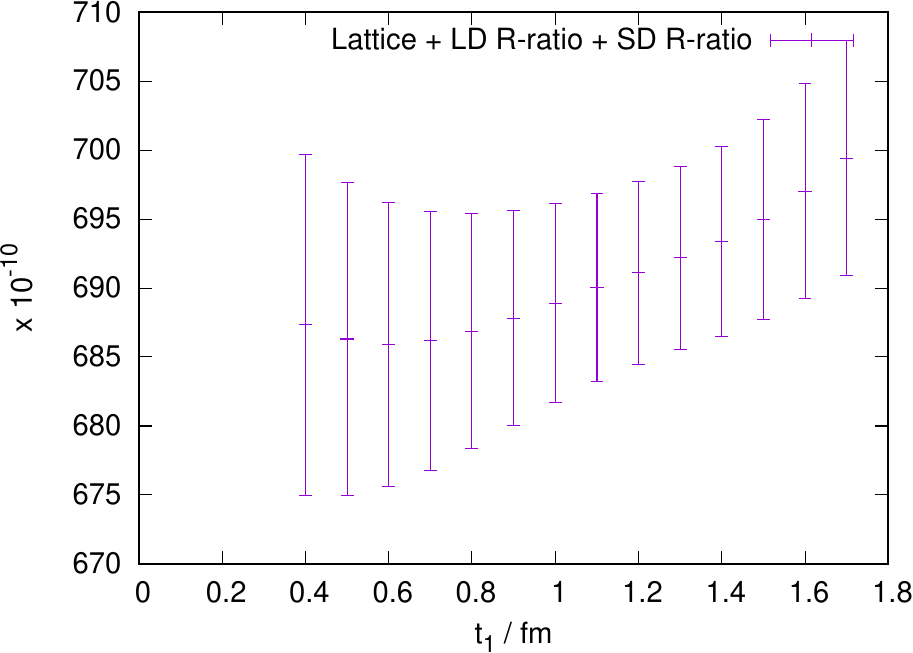}\\
            {\footnotesize sQED finite-volume correction}
          \end{center}
        \end{minipage}

       \vspace{0.3cm}

        \begin{minipage}{7cm}
          \begin{center}
            \includegraphics[width=5.5cm,page=16]{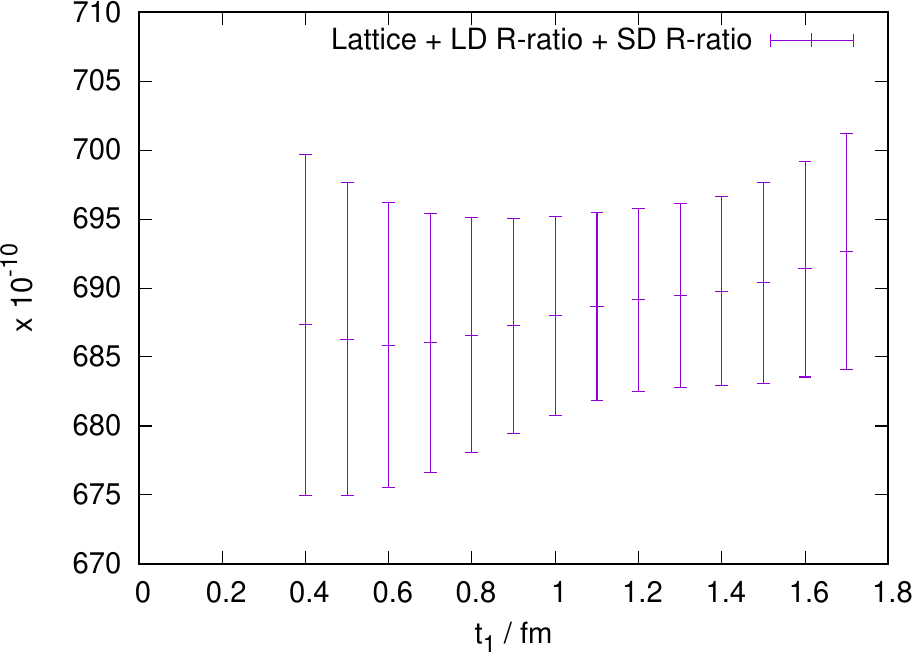}\\
            {\footnotesize quark-disconnected contribution}
          \end{center}
        \end{minipage}
       \begin{minipage}{7cm}
         \begin{center}
           \includegraphics[width=5.5cm,page=17]{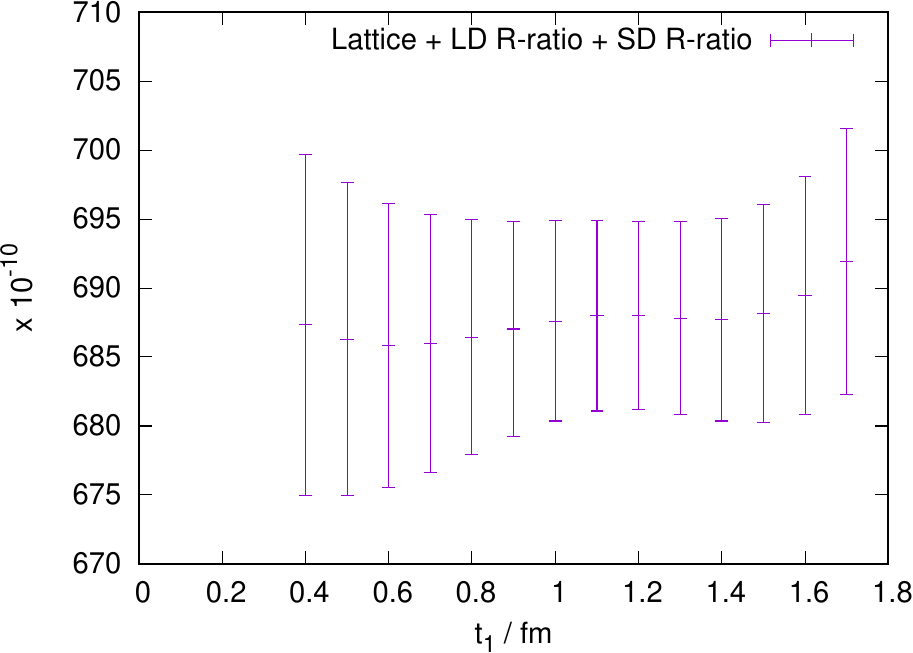}\\
           {\footnotesize Diagram M}
         \end{center}
       \end{minipage}
       
       \vspace{0.3cm}
       
       \begin{minipage}{7cm}
         \begin{center}
           \includegraphics[width=5.5cm,page=18]{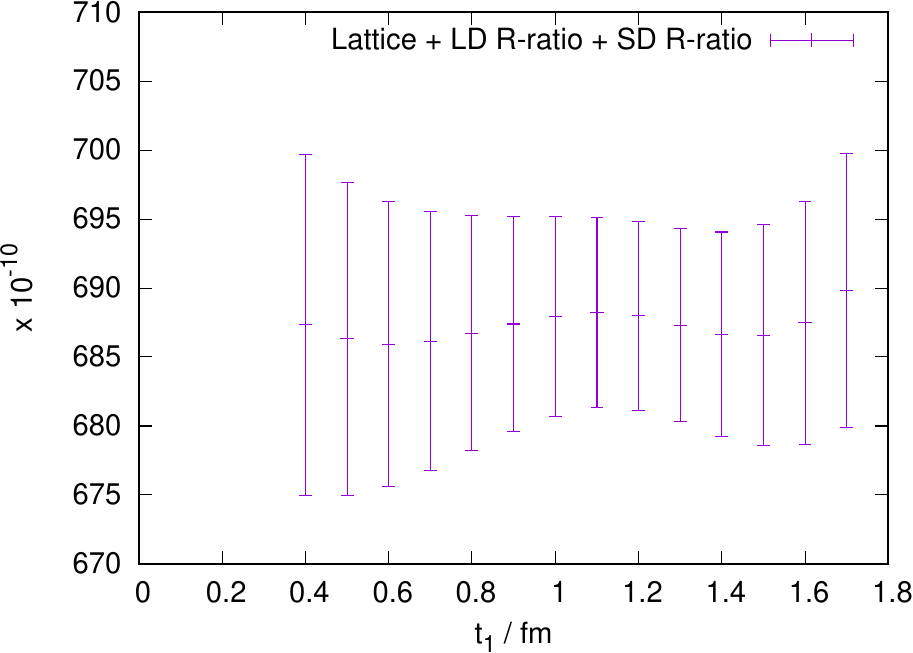}\\
           {\footnotesize Diagrams S+V}
         \end{center}
       \end{minipage}
       \begin{minipage}{7cm}
         \begin{center}
           \includegraphics[width=5.5cm,page=19]{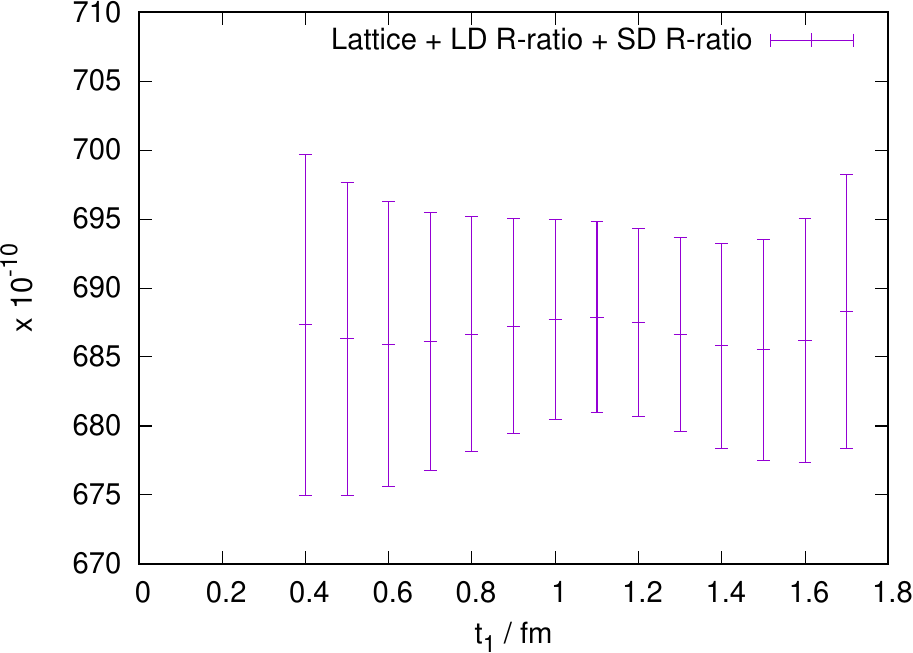}\\
           {\footnotesize Diagram F}
         \end{center}
       \end{minipage}

       \vspace{0.3cm}

       \begin{minipage}{7cm}
         \begin{center}
         \includegraphics[width=5.5cm,page=20]{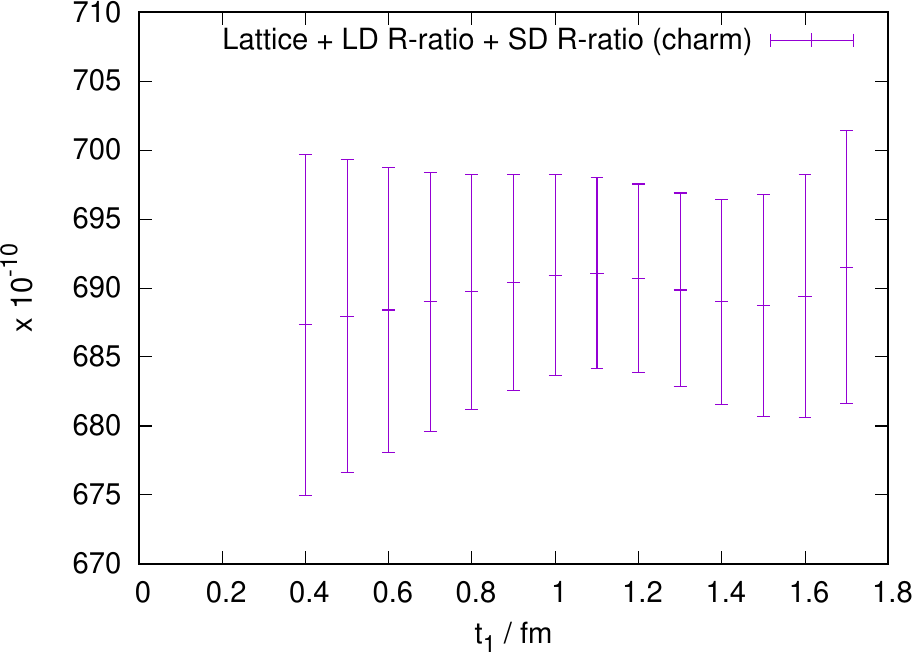}\\
         {\footnotesize charm quark-connected contribution}
         \end{center}
       \end{minipage}

      %\end{center}
       \caption{Effect of individual corrections to the
         isospin-symmetric quark-connected light and strange quark
         lattice data.  Contributions are accumulated from
         the top left to the bottom right labeling each new contribution.}
      % (\cite{\tiny DeGrand \& Sch\"afer 2004})
      \label{fig-m9f}% Give a unique label
    \end{figure}

In Fig.~\ref{fig-m9e} we show $a_\mu^{\rm LD}$ and $a_\mu^{\rm W}$
both individually as well as their sum for fixed $t_0=0.4$ fm and
$\Delta=0.15$ fm as a function of $t_1$.  In Fig.~\ref{fig-m9e} only
the isospin symmetric light and strange quark-connected data is used
for the lattice data.  In Fig.~\ref{fig-m9f}, we show the quantitative
effect of additional corrections to the lattice data: i) the addition
of scalar QED finite-volume corrections, ii) the addition of
quark-disconnected contributions, iii) the addition of strong
isospin-breaking diagram M, iv) the addition of QED corrections S+V,
v) the addition of QED correction F, and finally vi) the addition of
the quark-connected charm-quark contribution.

We observe that by changing $t_1$ from $0.4$ fm to $1.7$ fm we can
vary the fractional individual contributions from a region where the
result is given completely by the R-ratio data, to a region where the
result is mostly given by the lattice data.  In this range the
individual contributions therefore change on the order of $a_\mu^{\rm
  HVP~LO}$.  We note that the sum of both contributions is constant
within errors, which is a strong test of consistency between lattice
and R-ratio data.  As the lattice data becomes more precise in the
future, the total uncertainty can be reduced and longer distance
contributions can be calculated from ab-initio lattice computations.
In an upcoming paper \cite{Lehner:2017}, we also present results with
$a_\mu^{\rm SD}$ and $a_\mu^{\rm LD}$ obtained from the lattice
directly.

%      \begin{center}
%        \includegraphics[width=10cm,page=10]{figs/comb}
%      \end{center}

\section{Conclusion}
I have presented the current status of a first-principles calculation
at physical pion mass of the quark-connected, quark-disconnected,
leading QED and strong isospin-breaking contributions to the
leading-order HVP performed by the RBC and UKQCD collaborations.

The light-quark quark-connected contribution is computed with an
improved statistical estimator that reduces the statistical noise at
same cost compared to the usual $Z_2$ wall source method by an order of magnitude.
In addition a new Lanczos method using multiple grids \cite{Lehner:2017lanc} further improves
the efficiency of this method by reducing the memory cost by an order
of magnitude.  We combine this new dataset with the strange-quark
quark-connected and the quark-disconnected contributions
already published in Refs.~\cite{Blum:2016xpd} and
\cite{Blum:2015you}.

For the QED contributions we used a stochastic position-space sampling technique
similar to the one used in our hadronic light-by-light calculation of Ref.~\cite{Blum:2015gfa}.
We have re-tuned the quark masses and checked the prediction of the pion mass
splitting against the experimental result.  The re-tuning of the lattice spacing
and the QED corrections to the local current normalization are missing in this
presentation but will be published in Ref.~\cite{Lehner:2017}.

For the charm-quark contribution we use the vector correlator
of a heavy quark tuned by matching the heavy-light pseudo-scalar 
meson mass to the $D$ meson mass.

    \begin{figure}[tbp] % no figure before 1st section
      \begin{center}
        \includegraphics[width=9cm,page=23]{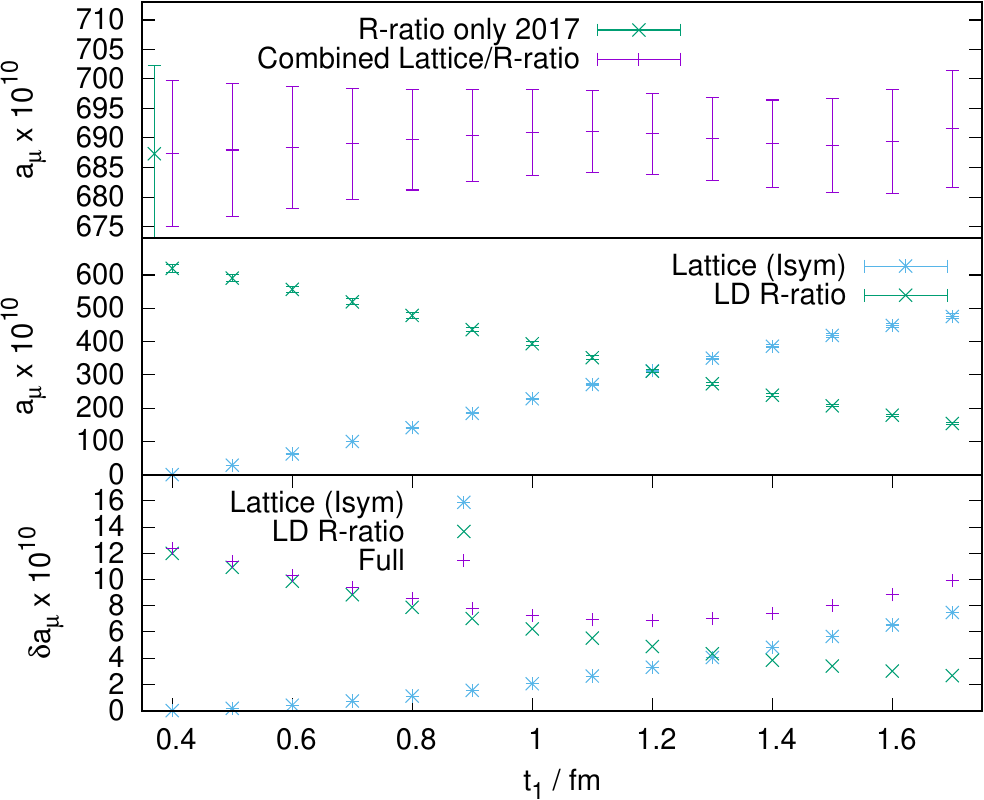}
      \end{center}
      \caption{The top panel shows the combined lattice and R-ratio
        result as a function of $t_1$.  We also show a pure R-ratio
        result based on the dataset of \cite{alphaQED16} ignoring the
        correlations between different energies.  The middle panel
        separates the size of the lattice and R-ratio contributions
        and the bottom panel shows the magnitude of individual
        contributions to the combined uncertainty.}
      \label{fig-m9g}% Give a unique label
    \end{figure}

We have devised a window method to combine and cross-check lattice and R-ratio data.  We summarize
the results again in Fig.~\ref{fig-m9g}.
This method allows for further reduction in uncertainty over the already very precise R-ratio results.
In the results presented here, correlations between different energies in the R-ratio are not yet
taken into account, however, this imperfection will be removed in Ref.~\cite{Lehner:2017}.

With these limitations we obtained a result with total uncertainty of
$\delta a_\mu^{\rm HVP~LO}=6.8 \times 10^{-10}$.  A full discussion of
systematic uncertainties will be provided in Ref.~\cite{Lehner:2017}.

\begin{figure}[tbp] % no figure before 1st section
      \begin{center}
        \includegraphics[width=9cm,page=2]{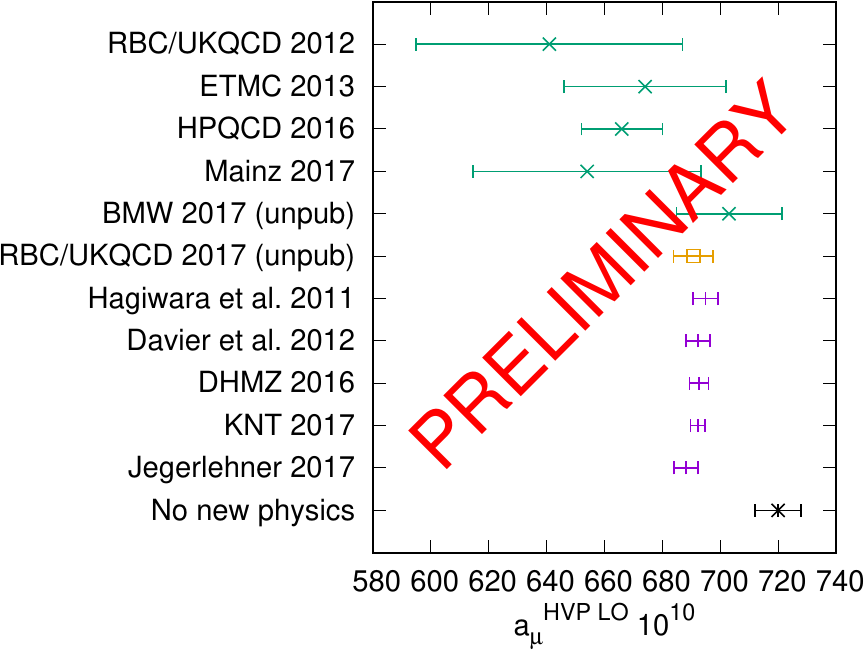}
      \end{center}
      \caption{Overview over recent lattice and R-ratio results.}
      % (\cite{\tiny DeGrand \& Sch\"afer 2004})
      \label{fig-m9h}% Give a unique label
    \end{figure}

We note that the window can eventually be widened to obtain a pure lattice
result with adequate precision matching the Fermilab E989 experiment.  In the near future,
however, the combination of R-ratio data with lattice data has the potential to
provide the most precise determination of the HVP.  We conclude
with showing our preliminary result in context of recent pure lattice and R-ratio
determinations in Fig.~\ref{fig-m9h}.

    {\bf Acknowledgments.}  The presented work is performed in
    collaboration with colleagues of the RBC and UKQCD collaborations.
    We are also indebted to Kim Maltman for a fruitful collaboration.
    C.L.~acknowledges support through a DOE Office of Science Early
    Career Award and by US DOE Contract DESC0012704(BNL).
    Computational resources provided by the clusters operated by USQCD
    at Fermilab and Jefferson Laboratory, by the Dirac facility at the
    University of Edinburgh, by the KNL system operated by the CSI at
    Brookhaven National Laboratory, and by the Mira facility at
    Argonne National Laboratory are gratefully acknowledged.

\clearpage
\bibliography{lattice2017}

%%%%%%%%%%%%%%%%%%%%%%%%%%%%%%%%%%%%%%%%%%%%%%%%%%%%%%%%%%%%%%%%%%%%%%%%%%%%%
\end{document}